\documentclass{aastex}

\usepackage{url}
\usepackage{graphicx,epsfig}
\usepackage{amsmath}
\RequirePackage{color}

\def\spose#1{\hbox to 0pt{#1\hss}}
\def\lta{\mathrel{\spose{\lower 3pt\hbox{$\mathchar"218$}}
     \raise 2.0pt\hbox{$\mathchar"13C$}}}
\def\gta{\mathrel{\spose{\lower 3pt\hbox{$\mathchar"218$}}
     \raise 2.0pt\hbox{$\mathchar"13E$}}}

\begin{document}

\setlength{\overfullrule}{0pt}

\title{Evolution of solar-type activity: \\
acoustic and magnetic energy generation and propagation \\
in $\beta$~Hydri (G2~IV)
}

\shorttitle{Acoustic and magnetic energy in Beta Hydri}
\shortauthors{Fawzy \& Cuntz}

\author{Diaa E. Fawzy\altaffilmark{1}} \and \author{M. Cuntz\altaffilmark{2}}

\altaffiltext{1}{Faculty of Engineering, Izmir University of Economics,
       35330 Izmir, Turkey.\\
       email: diaa.gadelmavla@ieu.edu.tr
}
\altaffiltext{2}{Department of Physics, University of Texas at Arlington, Box 19059,
       Arlington, TX 76019, USA.\\
       email: cuntz@uta.edu
}

\begin{abstract}
We examine the acoustic and magnetic energy generation and propagation in $\beta$~Hydri
(G2~IV).  The underlying motivation for this work is based on the solar, stellar, and
galactic relevance of $\beta$~Hyi (a star in the Southern hemisphere), which is
readily understood as a prime example and proxy of the future Sun --- thus allowing
assessments and analyses of the secular decay of solar activity.  Regarding the
magnetic energy generation, we consider longitudinal flux tube waves.  We also
assess acoustic waves.  For the acoustic wave energy flux, the difference between
the results obtained for $\beta$~Hyi and the Sun is significantly smaller than
typically attained for main-sequence stars, which is largely due to the
gravity-dependence of the acoustic energy generation.  Furthermore, we study
the height-dependent behavior of the magnetic energy flux for different magnetic
filling factors corresponding to different flux tube spreadings.
Finally, we comment on possible directions of future research.
\end{abstract}


\section{Introduction}

Solar-type stars, while situated on the main-sequence, are known to
exhibit fundamental changes over long timescales, i.e., millions and
billions of years.  The changes pertain both the stellar interior as
well as the atmospheres and winds, entailing large sets of observational
consequences.  In this study, we focus on $\beta$~Hydri
($\beta$~Hyi; HD 2151, HR 98, HIP 2021), a star in the Southern Hemisphere
that is part of the constellation Hydrus.  Previous modelling by \cite{bran11}
has reconfirmed $\beta$~Hyi's spectral type as G2~IV, among other spectral
properties.

\cite{drav93a,drav93b,drav93c} noted various aspects of interests about
$\beta$~Hyi, which are
(1) intriguing facts about its photosphere such as relatively large (a factor of
$\simeq$~5) convective cells (granules) as revealed by 3-D numerical simulations,
(2) reduced photospheric pressure leading to higher granular velocities,
(3) decreased chromospheric activity compared to the Sun (but showing intricate time variability),
(4) fainter transition region and coronal emission relative to the Sun, and
(5) stellar wind conditions shaped by post--main-sequence thermodynamics.
Based on further work, $\beta$~Hyi's age has been identified as about
6.5~Gyr \citep{drav98,bran11}.

Another aspect of $\beta$~Hyi is the existence of solar-type oscillations
\citep[e.g.,][]{bedd01,carr01,bedd07,fern03,maur03,doga10}, which allowed to further
establish $\beta$~Hyi's evolutionary status and to place additional constraints
on $\beta$~Hyi's stellar parameters.  The asteroseismic analysis of those data
allowed to identify both differences and similarities to the solar spectrum.
In addition, those detailed studies of $\beta$~Hyi served as an important
application and test-bed of existing stellar evolution and structure codes.

Stars akin to $\beta$~Hyi are characterized by reduced outer atmospheric
heating as well as a noticeably decreased level of chromospheric emission,
with the latter confirmed by observations \citep[e.g.,][]{drav93b,bucc08}.
Clearly, when stars evolve away from the main-sequence approaching the
subgiant stage, the magnitude of magnetic activity is considerably reduced
as a consequence of the restructuring of the stellar interior, associated
with the redistribution of the angular momentum, and the loss of angular
momentum due to the stellar wind \citep[e.g.,][]{schr93,char97,john15}; see also
\cite{metc22} for updated analyses.  Preliminary results for $\beta$~Hyi
based on theoretical two-component chromosphere models (acoustic and magnetic)
have been given by \cite{cunt22}, which also include some comparisons with
observations.

In this work, we explore aspects of acoustic and magnetic energy generation
in $\beta$~Hyi with a focus on longitudinal flux-tube waves (LTWs) and to
a lesser degree on acoustic waves (ACWs).  In Sect.~2, we convey our
theoretical approach, including a discussion of the stellar parameters
and the employed flux-tube models.  Our results and discussion are
given in Sect.~3.  In Sect.~4, we present our summary and conclusions.


\section{Theoretical approach}

\subsection{Stellar parameters}


Studies of acoustic and magnetic energy generation require the usage
of various stellar parameters.  In case of $\beta$~Hyi information is
given in Table~1.  As pointed out in Sect. 2.2., the initial acoustic
and magnetic wave energy depend on the effective temperature $T_{\rm eff}$,
surface gravity ${\log}~g_\star$, and the metallicity.  The key stellar
parameters of $\beta$~Hyi read as follows: $M_\star=1.08~M_\odot$,
$R_\star=1.81~R_\odot$, $T_{\rm eff}=5872$~K, $L_\star=3.49~L_\odot$,
${\log}~g_\star=4.02$~(cgs), and ${\lbrack}{\rm Fe/H}{\rbrack}={-0.10}$
\citep{silv06,nort07,brun10,bran11}.

\cite{drav93a} thoroughly discussed the status of $\beta$~Hyi as a
nearby subgiant and a valid proxy for the future Sun, exhibiting greater
granular velocities (about a factor of $\simeq$1.5 to 2) noting that an
almost identical surface energy flux must be carried by a lower
photospheric gas density.  Moreover, the scale of the photospheric
granulation is expected to be larger, and a considerably smaller number
of granules will simultaneously fit on its stellar surface; an outcome that
is relevant for $\beta$~Hyi's presumed magnetic surface structure.

Furthermore, consensus has been achieved about
$\beta$~Hyi's effective temperature (see Table~2).  The age of
$\beta$~Hyi has been given as  about 6.5~Gyr \citep{drav98,bran11};
see Table~3.  \cite{bran11} applied detailed asteroseismic modelling, stellar
surface structure assessments, and uncertainty analysis to arrive at that
value.  Previous studies in part also based on $\beta$~Hyi's solar-like
oscillations have been given by \cite{carr01} and \cite{fern03}.  The
latter authors also focused on a determination of $\beta$~Hyi's mass,
providing further support of the interpretation that this star very closely
represents the future Sun.


\subsection{Theory of acoustic and magnetic energy generation}


Akin to our previous work, the amount of energy generated by
ACWs or LTWs in the stellar convective zone uses the concept
of the mixing-length theory, including adjustments made based
on previous 3-D (magneto-)hydrodynamic models of stellar
convection; see \cite{tram97} and \cite{stein09a,stein09b}.
Previous results in that regard have been given by
\cite{musi89,musi94,musi95}, \cite{ulms96}, and \cite{ulms01}.

For example, the work of \cite{ulms01} examines LTW wave energy
fluxes carried along vertical magnetic flux tubes embedded in
the atmospheres of late-type stars, including solar-type stars,
subgiants and giants.  In their work, the main physical process
responsible for the generation of these waves is the nonlinear
time-dependent response of the flux tubes to continuous and
impulsive external turbulent pressure fluctuations; the latter
are represented by an extended Kolmogorov spatial and modified
Gaussian temporal energy spectrum.  Underlying theoretical studies
about stellar convection include work by \cite{stef89},
\cite{catt91}, and \cite{nord97}.

In this study, we assume a mixing-length parameter of
$\alpha_{\rm ML} = 1.8$ as advocated by \cite{stein09a,stein09b}.
Our main magnetic models are based on $\eta$ = 0.85 (see definition below).
Figure 1 conveys the LTW energy spectra for $\beta$~Hyi; see Table~1
for information on the stellar parameters.  In fact, we present
both the not-smoothed version and the smoothed version of the data,
with the latter being utilized for the acoustic and magnetic
energy generation; see Sect. 3.1.


\subsection{Flux tubes models}


Another important step of our approach is the construction of unperturbed and
unheated initial magnetic flux-tube models; they are
characterized by outwardly decreasing temperatures.
This step requires specifying four physical parameters, which are: the stellar effective
temperature $T_{\rm eff}$, the surface gravity ${\log}~g_\star$, the magnetic field strength at
the base of the flux tube $B_{0} (z=0~{\rm km})$, and the magnetic filling factor $f$.
Another parameter (with limited influence on our results) is the stellar metallicity,
which for $\beta$~Hyi is very close to the solar value \citep{brun10}.
For more detailed discussions about the selected approach to call attention to the
work by \cite{ulms01} and references therein.  For the current computations, we consider
the thin flux-tube approximation \citep[e.g.,][]{robe79,spru81,ferr89,hasa03},
which is readily applicable to most chromospheric layers of late-type stars
\citep[e.g.,][]{sten78,sola93,yell09}.

The height dependence of the tube geometry is physically governed by the initial photospheric
magnetic filling factor and by the conservation of the horizonal magnetic pressure with the
external gas pressure, noting that the guiding equation is given as:
$B_{\rm eq}^2/8\pi \ + p_i = p_e$ with $p_{i}$ and $p_{e}$ being the internal and external gas
pressures at a given height.  The maximum tube opening is determined from the following equation:
$f = r^{2}_{\rm bottom} / r^{2}_{\rm top}$ where $r_{\rm bottom}$ and $r_{\rm top}$ are the bottom
and top radii  of the flux tube at the stellar surface and at the maximum opening height, respectively.
Akin to previous solar models \citep[e.g.,][]{fawz98}, the bottom radius of the magnetic flux tube is
assumed as about half the local pressure scale height.  For early studies of $\beta$~Hyi's
photospheric structure see \cite{drav93a}.

For the determination of the magnetic field strength at height $z = 0$~km, we first
compute the maximum allowed field of a void tube assuming the inner gas pressure $p_{i} = 0$
dyne~cm$^{-2}$; this values also known as the equipartition field strength given by
$\sqrt{p_{e} / 8\pi} = B_{\rm eq}$ taken at 0~km, the photospheric reference height.
According to the initial atmospheric model, the resulting equipartition magnetic field strength
is given as $B_{\rm eq} = 1505$~G.  A comparison to the solar case, as well as in alignment to
previous magneto-acoustic model simulations for other stars
\citep[e.g.,][]{fawz18,fawz21,cunt21}, we choose a surface magnetic field strength given as
$0.85~B_{\rm eq}$.  Nevertheless, as part of our parameter study of magnetic energy generation
based on longitudinal flux-tube waves, we consider a larger set of parameters $\eta = B/B_{\rm eq}$,
given as 0.75, 0.85, and 0.95.

In the current study, we consider three different surface magnetic filling factors, namely,
1\%, 5\%, and 10\%, resulting in a bottom radius of about $r_{\rm bottom} = 196$~km and
three top opening radii, given as $r_{\rm top} = 1750$~km, 781~km, and 552~km, respectively;
see Fig.~2.  Note that our work is most relevant for the low and middle part of $\beta$~Hyi's
chromosphere.  Future studies aimed at $\beta$~Hyi's high chromosphere and wind would need
to consider also the impact of wave coupling effects and wave mode conversions;
see, e.g., \cite{sriv21} and references therein.


\section{Results and discussion}

\subsection{Acoustic and magnetic energy generation}



The main component of this study concerns the acoustic and magnetic energy generation,
with a focus on LTWs, for $\beta$~Hyi as facilitated at photospheric levels.  As discussed
in Sect.~2.2, we make use of previous works in the literature \citep[e.g.,][]{ulms01}.
Regarding LTWs, these kinds of models consider the nonlinear time-dependent response
to external pressure fluctuations acting on the flux tubes embedded in the stellar atmosphere.
Following detailed 3-D simulations for the Sun, \cite{stein09a,stein09b}
indicated that $\alpha_{\rm ML} = 1.8$ might be most realistic.  Since it is still
unsettled how this kind of result would translate to stars of lower surface gravity,
calculations assuming a larger set of $\alpha_{\rm ML}$ are desirable.

Figure 3 depicts acoustic wave energy fluxes for mixing-length parameters between
1.6, 1.8, and 2.0.  For comparison, we also give the result for the solar case with
$\alpha_{\rm ML}$ = 1.8; see \cite{fawz11} and references therein for information
on the relevant model parameters.  Furthermore, Fig.~4 depicts a comparison
between the generated flux for LTWs for $\beta$~Hyi with mixing-length parameters
$\alpha_{\rm ML}$ = 1.6, 1.8, and 2.0 and for $\eta$ = 0.85.  Moreover, a comparison
between the results for $\beta$~Hyi pertaining to LTW fluxes for $\alpha_{\rm ML}$
= 1.8 and for $\eta$ = 0.75, 0.85, and 0.95 is conveyed in Fig.~5.  The numerical
values of the energy flux generation for ACWs and LTWs are given in Table 4 and 5,
respectively.

It is found that for ACWs, the amount of generated energy is notably higher (about a
factor of 2.3) in $\beta$~Hyi compared to the Sun.  This behavior closely related to
the greater photospheric granular velocity of $\beta$~Hyi due to its lower surface
gravity; see \cite{drav93a}.
However, for LTWs, the amount of generated energy is somewhat lower (about a
factor of 1.1 to 1.3) in $\beta$~Hyi compared to the Sun.  Hence, the difference
between magnetic and acoustic energy generation is relatively small for $\beta$~Hyi,
contrary to solar-type stars, including the Sun.
Additionally, it is found that lower values of $\eta = B/B_{\rm eq}$ lead to higher
amounts of magnetic energy generation, a finding akin to previous results for other stars,
especially main-sequence stars.  In fact, those results indicate that the
magnetic energy flux for tubes $\eta$ = 0.75 can be about one order of magnitude
higher compared to models of 0.95. This can be explained by the fact that the increase 
in $\eta$ increases the stiffness of the magnetic flux tube, which in turn
decreases the efficiency of the LTW wave generations.
 
Figure 6 depicts the two primary input wave energy spectra $\beta$~Hyi obtained in
this study.  We show the case of LTWs with  $\alpha_{\rm ML}$ = 1.8 and $\eta$ = 0.85
and the corresponding case of ACWs.  It is found that the difference in flux for LTWs
and ACWs is much smaller than readily obtained for main-sequence stars; in fact, the
wave energy flux for ACWs in evolved stars is relatively high owing to the
gravity-dependence of the acoustic energy generation \citep[e.g.,][]{ulms96}.
Specifically, for $\alpha_{\rm ML}$ = 1.8 and $\eta$ = 0.85 the ratio for the
wave energy fluxes between $\beta$~Hyi and the Sun reads 2.3 for ACWs and 0.87 for LTWs.
From an astrophysical perspective, the relatively high ACW energy flux in $\beta$~Hyi
is due to the relatively high convective flow speeds in subgiants relative to
main-sequence stars with comparable stellar parameters.


\subsection{Comments on wave heating models and flux tube spreading}


Figure 7 gives an example of a magnetic heating models for $\beta$~Hyi.
We convey a snapshot of a longitudinal flux tube model with MFF = 1\%
at an elapsed time of 3364~s.  Various quantities are depicted, including
a strong shock at height 2700 km with a shock strength measured as $M_{\rm sh}$ = 6.3.
Time-dependent ionization has been considered.  It is a monochromatic wave
model with a wave period of 400~s, a value significantly higher than in
models previously given for the Sun in responds to $\beta$~Hyi's reduced
surface gravity.  A total of 14 shocks have been inserted.
Additional time-dependent results for $\beta$~Hyi's magnetically heated
outer atmosphere have previously been given by \cite{cunt22}.
These  models also considered the computation of Ca~II fluxes,
including tentative comparisons with observational constraints.

In Figure 8, we examine the height-dependent behavior of the mechanical energy flux
for LTWs regarding $\beta$~Hyi.  We consider both monochromatic and spectral waves
while considering models based on MFF of 1\%, 5\%, and 10\%, respectively.
The initial wave energy fluxes are given as $2.74 \cdot 10^8$ erg~cm$^{-2}$~s$^{-1}$;
this value corresponds to our main model given as $\alpha_{\rm ML}$ = 1.8 and $\eta$ = 0.85.
It is found that a higher MFF entails a somewhat smaller decrease of the LTW
energy flux as a function of height, especially in the middle chromosphere,
mostly associated with the difference in tube spreadings.  However, at
large heights other effects are relevant as well, especially in narrow tubes,
including effects associated with strong shocks, which initiate both
quasi-adiabatic cooling and a reduction of the time-averaged densities;
see, e.g., \cite{fawz12} for previous results.


%
%
\begin{table}
\caption{Stellar Parameters}
\centerline{\begin{tabular}{l c l} \hline
\noalign{\smallskip}
Parameter & Value & Reference \\
\noalign{\smallskip}
\hline
\hline
\noalign{\smallskip}
$M$~($M_\odot$)        & 1.08  $\pm$ 0.03   & \cite{bran11}  \\
$R$~($R_\odot$)        & 1.809 $\pm$ 0.015  & \cite{bran11}  \\
$T_{\rm eff}$~(K)      & 5872  $\pm$ 44     & \cite{nort07}  \\
$L$~($L_\odot$)        & 3.494 $\pm$ 0.087  & \cite{bran11}  \\
${\log}~g_\star$~(cgs) & 4.02  $\pm$ 0.04   & \cite{silv06}  \\
{\lbrack}Fe/H{\rbrack} & $-$0.10 $\pm$ 0.07 & \cite{brun10}  \\
Age~(Gyr)              & 6.40 $\pm$ 0.56    & \cite{bran11}  \\
\hline
\noalign{\smallskip}
\end{tabular}
}
\noindent
Note: All symbols have their usual meaning.
\end{table}


%
%
\begin{table}
\caption{Determination of $T_{\rm eff}$}
\centerline{\begin{tabular}{l c l} \hline
\noalign{\smallskip}
Parameter & Value & Reference \\
\noalign{\smallskip}
\hline
\hline
\noalign{\smallskip}
$T_{\rm eff}$~(K)  & 5872 $\pm$ 44  & \cite{nort07}  \\
...                & 5774 $\pm$ 60  & \cite{bene98}  \\
...                & 5800 $\pm$ 100 & \cite{drav93a} \\
	\hline
\end{tabular}
}
\end{table}

%
%
\begin{table}
\caption{Determination of the Stellar Age}
\centerline{\begin{tabular}{l c l} \hline
\noalign{\smallskip}
Parameter & Value & Reference \\
\noalign{\smallskip}
\hline
\hline
\noalign{\smallskip}
Age~(Gyr) & 6.40 $\pm$ 0.56 & \cite{bran11}  \\
...       & 6.75 $\pm$ 0.35 & \cite{fern03}  \\
...       & $\sim$6.7       & \cite{drav98}  \\
...       & 9.5  $\pm$ 0.8  & \cite{drav93a} \\
\hline
\end{tabular}
}
\end{table}

%
%
\begin{table}
\caption{Acoustic Energy Generation}
\centerline{\begin{tabular}{l c c} \hline
\noalign{\smallskip}
$\alpha_{\rm ML}$ &  \multicolumn{2}{c}{Wave Energy Flux$^a$} \\
\noalign{\smallskip}
...               &  $\beta$~Hyi     &  Sun               \\
\noalign{\smallskip}
\hline
\hline
\noalign{\smallskip}
1.6  &  1.62e+8   &  6.94e+7 \\
1.8  &  2.51e+8   &  1.05e+8 \\
2.0  &  3.65e+8   &  1.58e+8 \\
\hline
\noalign{\smallskip}
\end{tabular}
}
\noindent
$^a$Unit: erg~cm$^{-2}$~s$^{-1}$
\end{table}

%
%
\begin{table}
\caption{LTW Energy Generation}
\centerline{\begin{tabular}{l c c c} \hline
\noalign{\smallskip}
$\alpha_{\rm ML}$ & $\eta$ & \multicolumn{2}{c}{Wave Energy Flux$^a$} \\
\noalign{\smallskip}
...               & ...    & $\beta$~Hyi     &  Sun               \\
\noalign{\smallskip}
\hline
\hline
\noalign{\smallskip}
1.6  &  0.75  &  3.65e+8  &  4.54e+8  \\
...  &  0.85  &  2.19e+8  &  2.55e+8  \\
...  &  0.95  &  6.47e+7  &  7.27e+7  \\
1.8  &  0.75  &  4.89e+8  &  5.52e+8  \\
...  &  0.85  &  2.74e+8  &  3.15e+8  \\
...  &  0.95  &  7.22e+7  &  9.41e+7  \\
2.0  &  0.75  &  5.79e+8  &  6.84e+8  \\
...  &  0.85  &  2.88e+8  &  3.99e+8  \\
...  &  0.95  &  8.92e+7  &  1.13e+8  \\
\hline
\noalign{\smallskip}
\end{tabular}
}
\noindent
$^a$Unit: erg~cm$^{-2}$~s$^{-1}$
\end{table}

%
\begin{figure*}[h!]
\hspace{0.5cm}
  \includegraphics[width=0.45\textwidth]{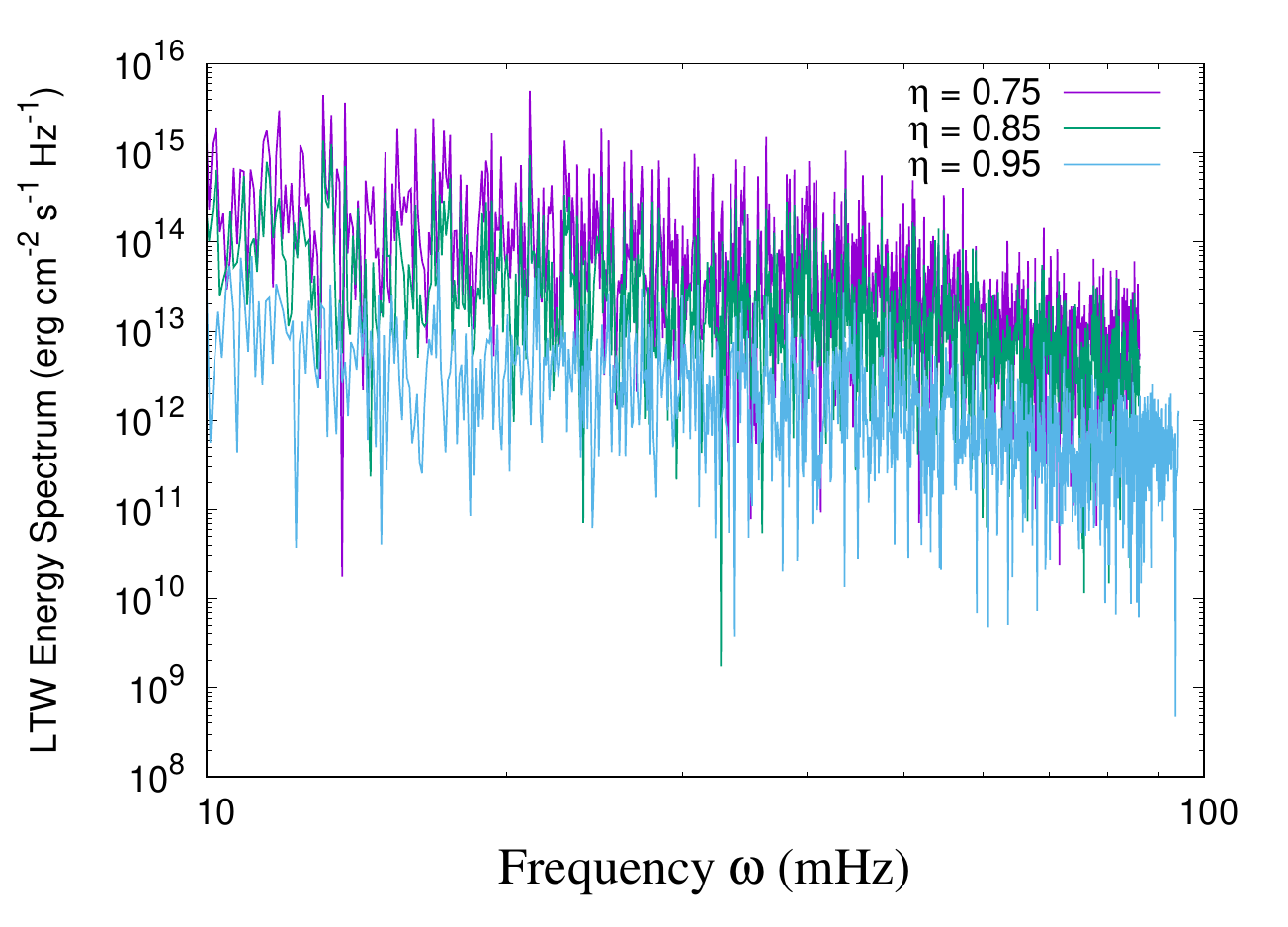}
  \includegraphics[width=0.45\textwidth]{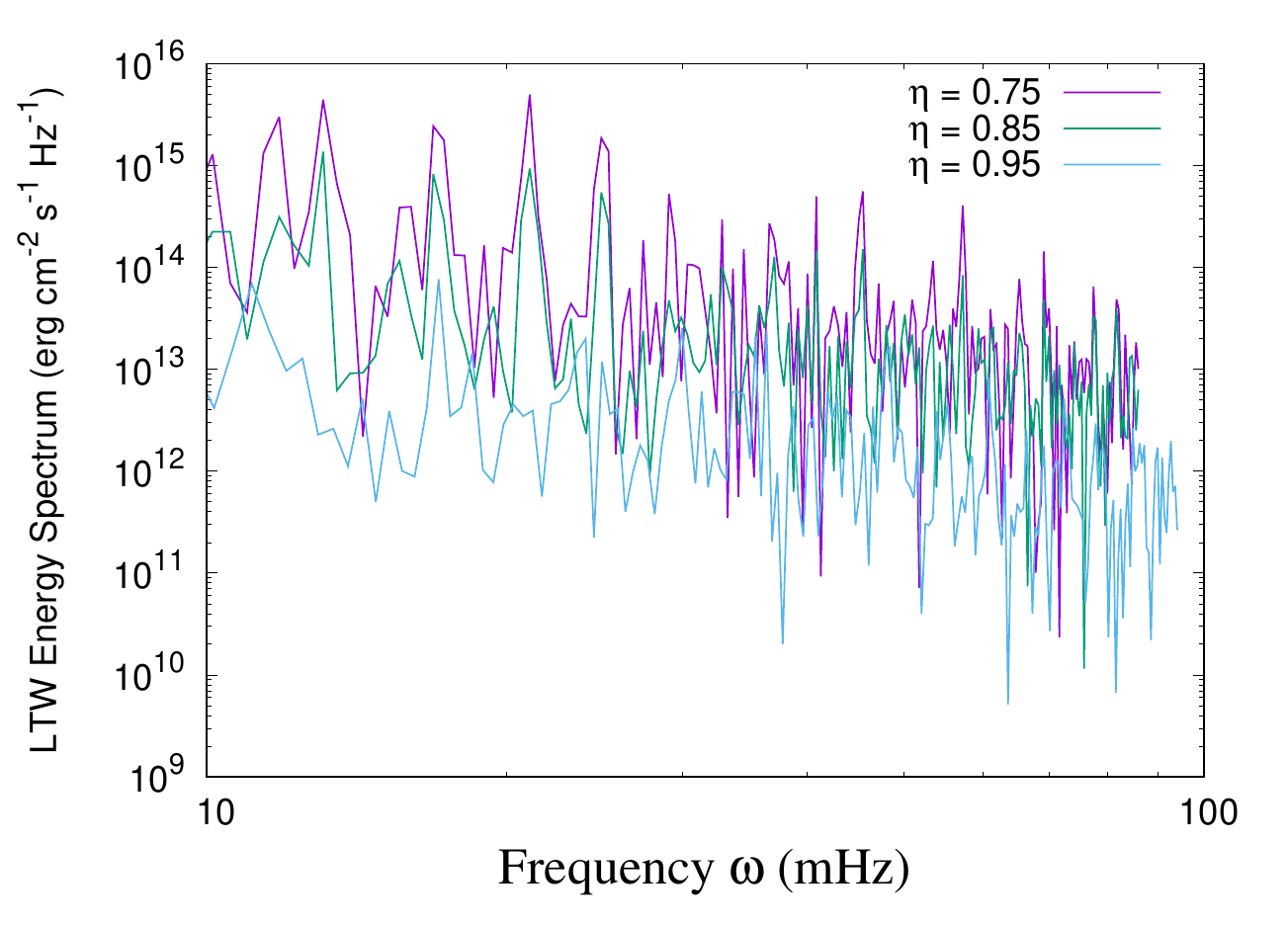}
     \caption{LTW energy spectra for $\beta$~Hyi assuming $\alpha_{\rm ML}$ = 1.8 and different values of $\eta$
              while depicting the not-smoothed version (left panel) and the smoothed version of the data (right panel).}
\end{figure*}

%
\begin{figure}[h!]
\hspace{1cm}
  \includegraphics[width=0.45\textwidth]{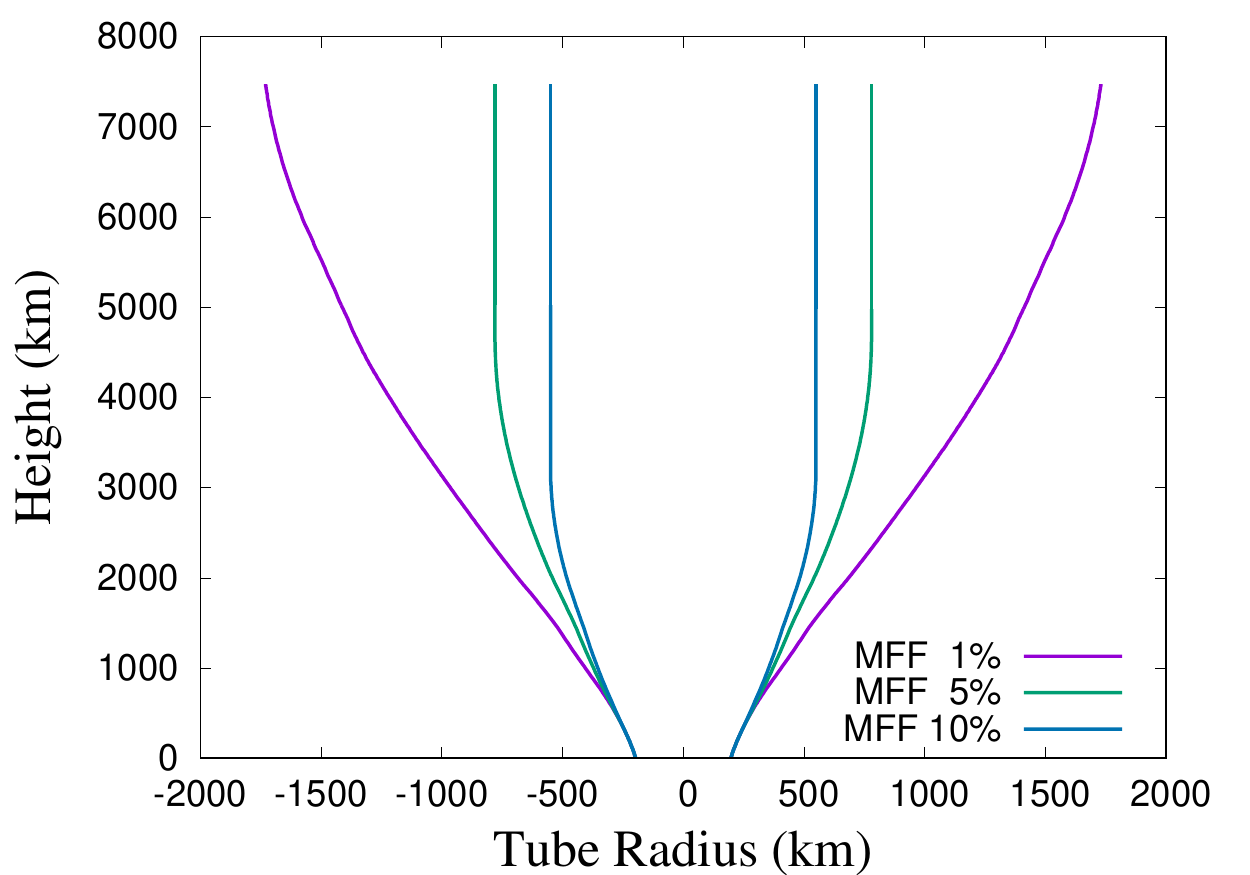}
     \caption{Flux tube models for $\beta$~Hyi regarding different magnetic filling factors.}
\end{figure}

%
\begin{figure}[h!]
\hspace{1cm}
  \includegraphics[width=0.45\textwidth]{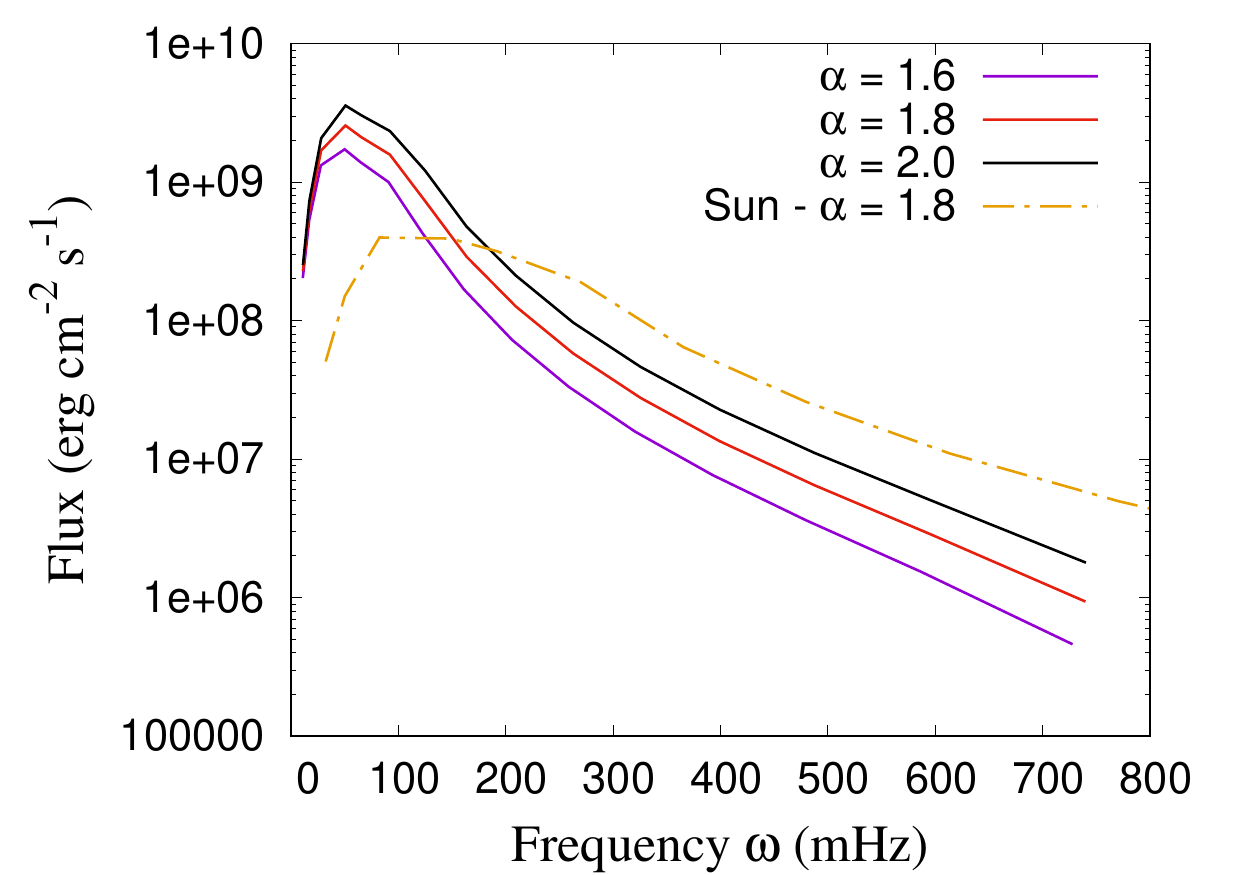}
     \caption{Acoustic wave energy fluxes for $\beta$~Hyi with mixing-length parameters
$\alpha_{\rm ML}$ = 1.6, 1.8, and 2.0 and the Sun assuming $\alpha_{\rm ML}$ = 1.8.}
\end{figure}

%
\begin{figure*}[h!]
\hspace{1cm}
  \includegraphics[width=0.85\textwidth]{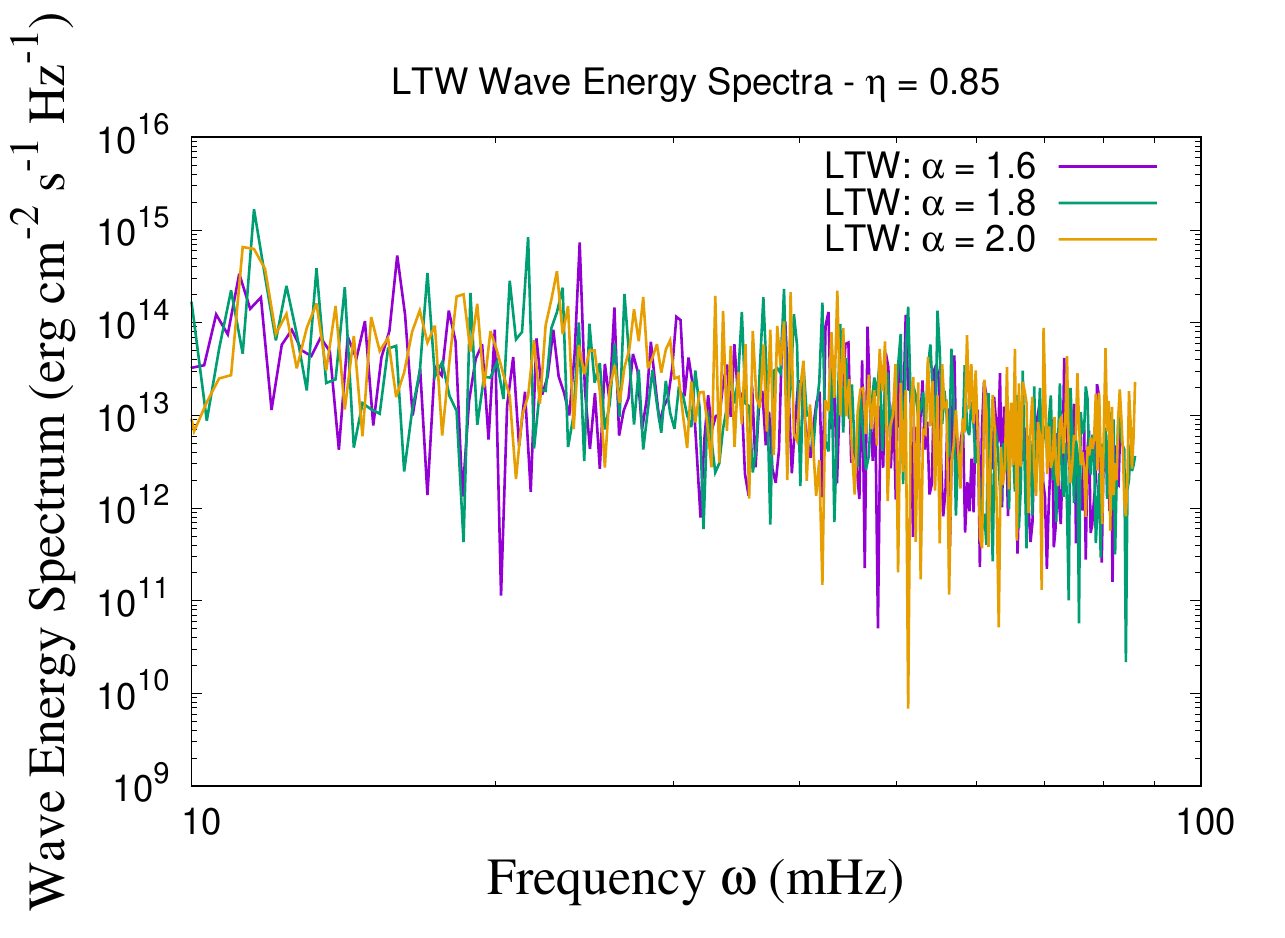}
     \caption{A comparison between the generated flux for LTWs for $\beta$~Hyi
with mixing-length parameters $\alpha_{\rm ML}$ = 1.6, 1.8, and 2.0 and
for $\eta$ = 0.85.}
\end{figure*}

%
\begin{figure*}[h!]
\hspace{0.5cm}
  \includegraphics[width=0.85\textwidth]{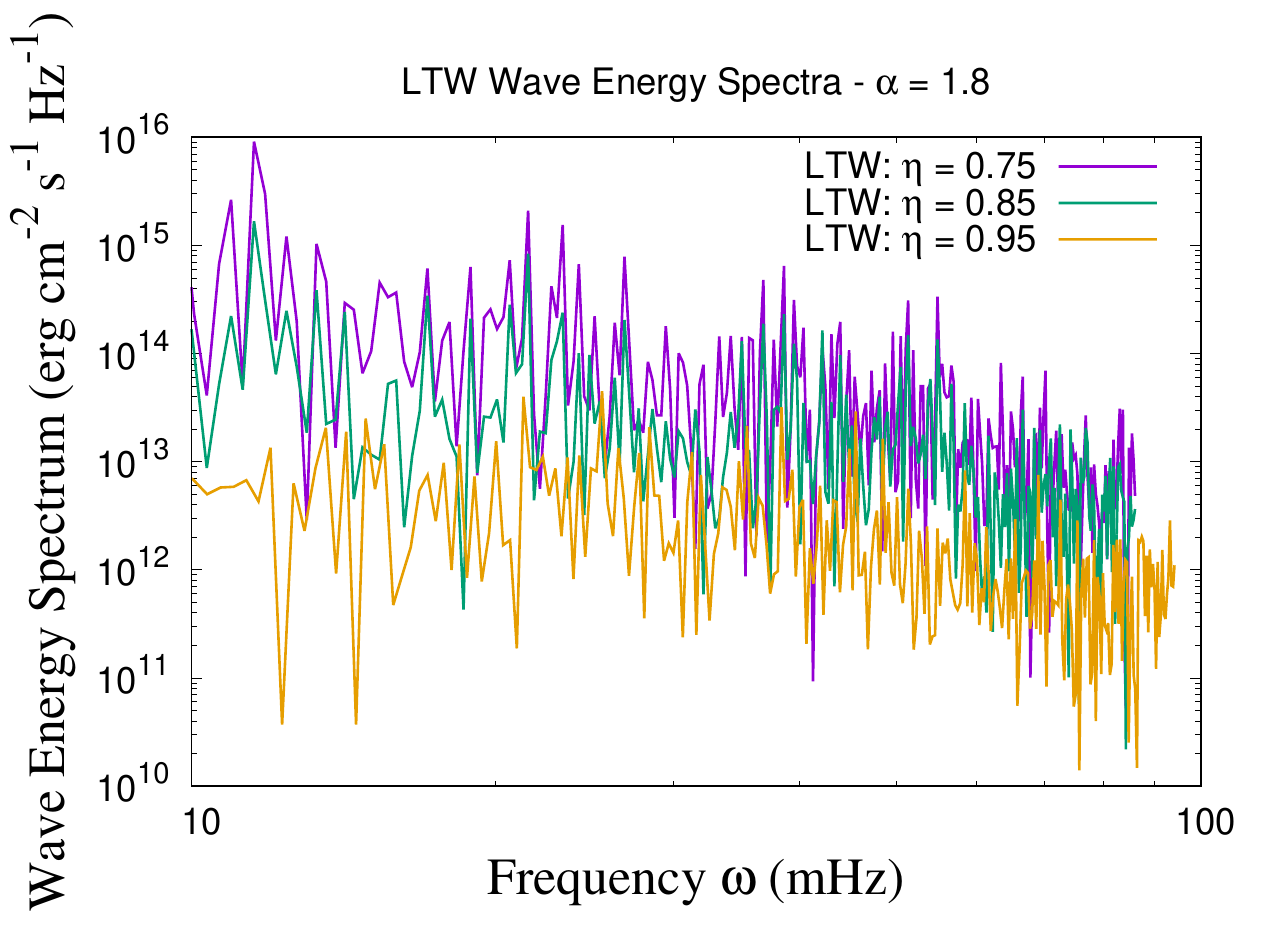}
     \caption{A comparison between the generated flux for LTWs for $\beta$~Hyi
with mixing-length parameter $\alpha_{\rm ML}$ = 1.8 and for $\eta$ = 0.75, 0.85,
and 0.95.}
\end{figure*}

%
\begin{figure}[h!]
\hspace{1cm}
  \includegraphics[width=0.45\textwidth]{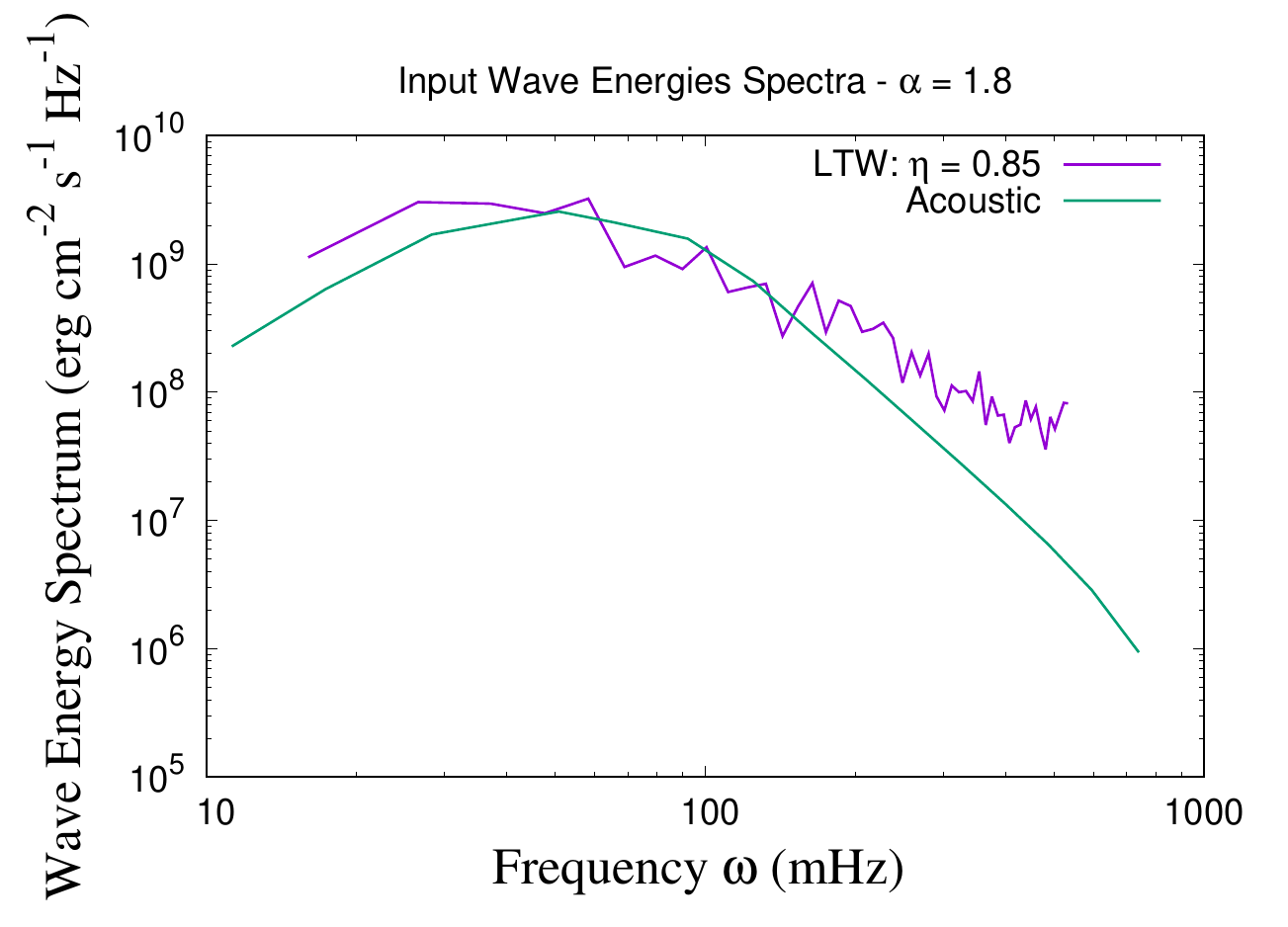}
     \caption{Depiction of input wave energy spectra $\beta$~Hyi.  We show the case of longitudinal flux-tube waves with
$\alpha_{\rm ML}$ = 1.8 and $\eta$ = 0.85 and the corresponding case of acoustic waves.}
\end{figure}

%
\begin{figure*}[h!]
\hspace{1cm}
  \includegraphics[width=0.95\textwidth]{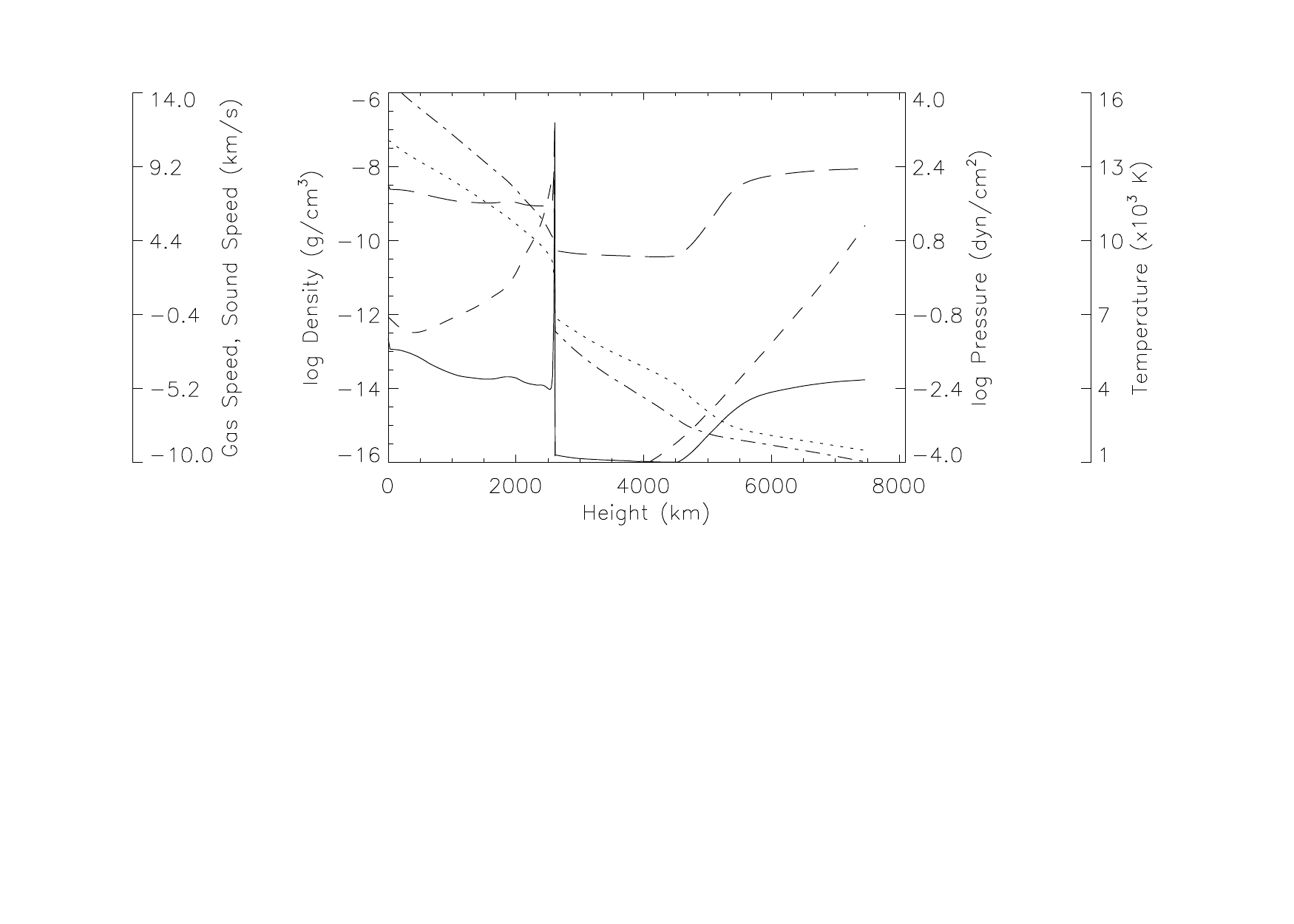}
     \caption{Snapshot of a longitudinal flux tube model with MFF = 1\%
at an elapsed time of 3364~s.  The following quantities are shown:
temperature (solid line), gas density (short-dashed line), gas speed (midway-dashed line),
sound speed (long-dashed line), and gas pressure (dashed-dotted line).}
\end{figure*}

%
\begin{figure*}[h!]
\hspace{1cm}
  \includegraphics[width=0.75\textwidth]{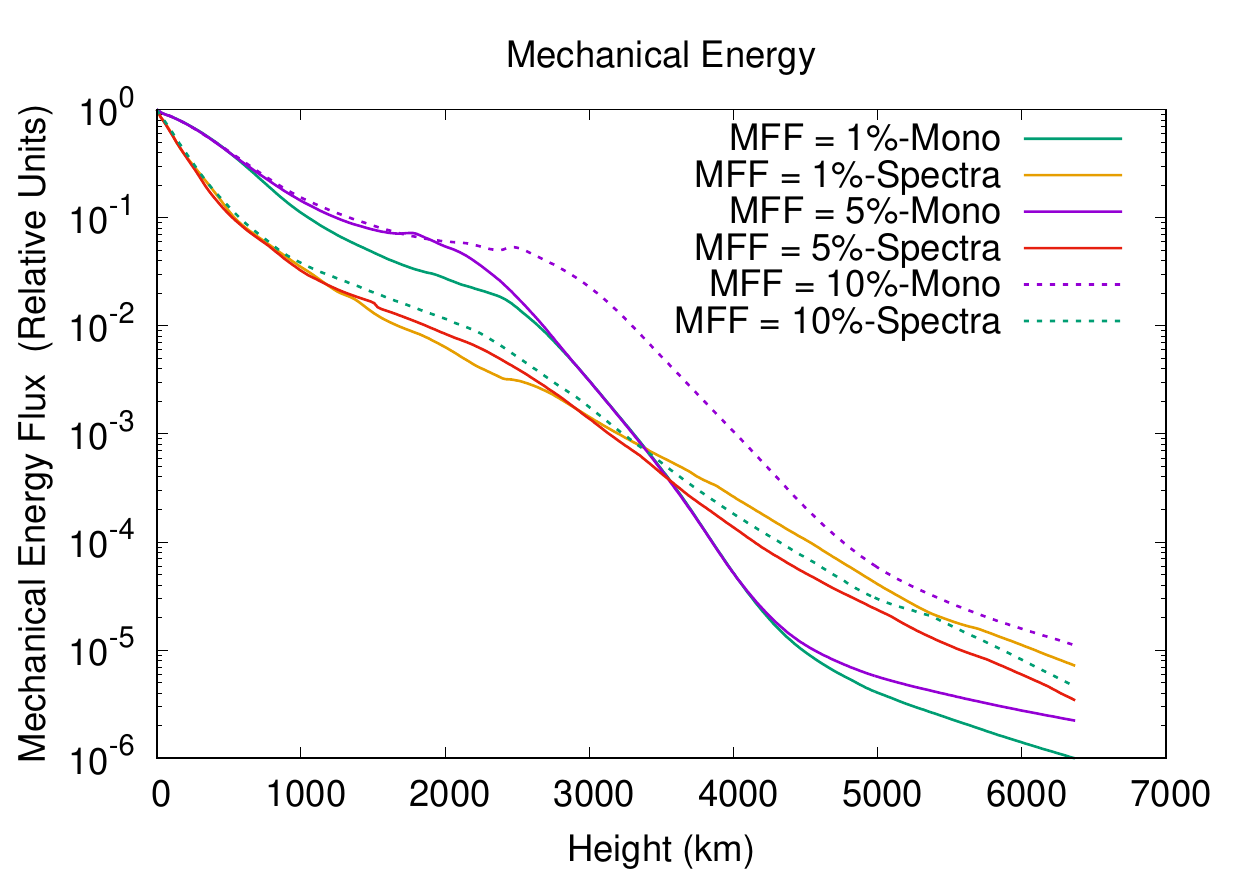}
     \caption{Height-dependent behavior of the mechanical energy flux in models of longitudinal flux-tube waves (relative units) for $\beta$~Hyi.
We consider both monochromatic waves and wave energy spectra assuming magnetic filling factors of 1\%, 5\%, and 10\%, respectively.
The initial wave energy fluxes are given as $2.74 \cdot 10^8$ erg~cm$^{-2}$~s$^{-1}$.}
\end{figure*}

\subsection{The big picture}

It is of obvious interest to compare both the acoustic and magnetic
heating in $\beta$~Hyi to the solar case.  Previous work by \cite{bohn84}
determined that the amount of the generated acoustic energy obeys the
relation $F_{\rm M}^{\rm AC} \propto T_{\rm eff}^{9.75} g_\star^{-0.5}$ with
$T_{\rm eff}$ and $g_\star$ as stellar effective temperature and surface
gravity, respectively.  Based on $T_{\rm eff} = 5872 \pm 44$~K for $\beta$~Hyi
\citep{nort07}, the generated acoustic energy is increased between 10\%
and 25\% depending on the adopted value of $T_{\rm eff}$.  However,
based on  $\beta$~Hyi's surface gravity, $F_{\rm M}^{\rm AC}$ is increased
by a factor of about 2.5 (see Table 4), as expected for
evolved stars.  Concerning the LTW models, the difference in the wave
energy generation between $\beta$~Hyi and the Sun is notable as well.
For the different values of $\alpha_{\rm ML}$ and $\eta$ (see Table~5),
the initial wave energy in $\beta$~Hyi relative to the Sun is increased by
a factor of 1.1 to 1.3, depending on the model.

For evolved stars, exhibiting very low chromospheric activity, also referred
to as chromospheric basal flux stars, previous observational work by, e.g.,
\cite{schr87} and \cite{rutt91} indicates that their emergent emission in Ca~II
and Mg~II is close to the values obtained for low-activity main-sequence stars;
for more information see \cite{schr95}.  This outcome has also been reproduced
by detailed theoretical heating models as given by, e.g., \cite{ulms89},
\cite{buch98}, \cite{cunt94}, and \cite{fawz02}.

This behavior
is due to the fact that evolved stars possess more extended radiative zones.
In those stars, the wave energy flux decreases much more rapidly as a function
of height compared to main-sequence stars; hence, the amount of energy
available in the Ca~II and Mg~II formation regions is largely insensitive to the
amount of initial wave energy and (by implication) to the stellar surface gravity.
Stars akin to $\beta$~Hyi are characterized by moderate amounts of magnetic
activity; hence, those atmospheric structures are more complex.  Consequently,
more detailed models are required, ideally encompassing different kinds
of magnetic heating modes and magnetic field configurations.

Another approach has been pursued by \cite{pere11}.  They considered a set of
cool G, K, and M giants and supergiants (with evolutionary stages significantly
beyond that of $\beta$~Hyi) and inspected their Mg~II {\it h} + {\it k} emissions.
They found that the results (when considering the various statistical uncertainties)
agree well with the assumption of nonmagnetic energy dissipation (supposedly
provided by acoustic waves) in the turbulent chromospheres.  This is a stark
motivation to further investigate the relationship between (magneto-)acoustic
heating and chromospheric emission for $\beta$~Hyi, an intermediate case between
stars of considerable magnetic activity and chromospheric basal flux stars.


\section{Summary and conclusions}

The aim of this work has been the examination of the generation and propagation of
acoustic and magnetic energy regarding $\beta$~Hyi.  This star is considered a benchmark
G2~IV star situated in the solar neighborhood.
Early studies about $\beta$~Hyi \citep{drav93a,drav93b,drav93c} focused on
stellar evolutionary aspects, photospheric structure, chromospheric activity
and variability as well as its transition region, corona, and stellar wind.
In the meantime, $\beta$~Hyi's age has been determined as approximately 6.5~Gyr
\citep[e.g.,][]{bran11}, thus confirming its status as a proxy of the future Sun,
thus establishing $\beta$~Hyi's relevance for studies of stellar evolution and
circumstellar habitability \citep[e.g.,][ and related work]{sack93,maur03,riba05}.

In this study, we focus on calculations of the initial magnetic and acoustic energy
generation with a concentration on longitudinal flux-tubes waves.  Previous theoretical
work in that context has been given by \cite{musi89,musi94,musi95} and \cite{ulms01}.
Specifically, we consider models with MFF of 1\%, 5\%, and 10\% to explore the
height-dependent behavior of the magnetic wave energy flux.  As previously identified
for models of other stars, including the Sun \citep[e.g.][]{fawz98}, as smaller valuer
MFF generally implies reduced spreading of the magnetic flux tubes at both
photospheric and chromospheric heights.

Concerning ACW and LTW photospheric energy generation, we found the following results:

\medskip\noindent
(1) Both for ACWs and LTWs, a larger mixing-length parameter $\alpha_{\rm ML}$
entails a higher amount of energy generation in agreement with previous results
\citep[e.g.,][, and related work]{ulms96}.

\medskip\noindent
(2) Regarding ACWs, the amount of generated energy is notably higher (about a
factor of 2.3) in $\beta$~Hyi compared to the Sun.  This behavior is due to the
greater photospheric granular velocity of $\beta$~Hyi due to its lower
surface gravity; see \cite{drav93a}.

\medskip\noindent
(3) Regarding LTWs, the amount of generated energy is somewhat lower (about a
factor of 1.1 to 1.3) in $\beta$~Hyi, a behavior associated with its different
thermodynamic and magnetic field conditions.  Consequently, the difference
between magnetic and acoustic energy generation is relatively small for
$\beta$~Hyi, contrary to solar-type main-sequence stars.

\medskip\noindent
(4) Lower values of $\eta = B/B_{\rm eq}$ result in higher amounts of magnetic
energy generation, a finding akin to previous results for other stars, including
main-sequence stars; e.g., \cite{ulms01}.  In these stars the flux
can be about one order of magnitude higher in tubes with $\eta$ = 0.75 compared
to 0.95.

\medskip

The fact that both the acoustic and magnetic energy generation in
$\beta$~Hyi is reduced compared to the Sun is a consequence of
differences regarding their thermodynamic and magnetic properties
at photospheric heights.  In case of $\beta$~Hyi a reduced surface
density as well as a higher photospheric granular velocity
(convective speed) occurs.  Moreover, a higher MFF entails a somewhat
smaller decrease of the LTW energy flux as a function of height,
especially in the middle chromosphere, due to the MFF-dependent
tube spreadings.  At large heights, those differences are less pronounced
because the relatively high energy fluxes in narrow tubes initiate strong
shocks, which notably affect the thermal structure of the tubes due to
quasi-adiabatic cooling; see, e.g., \cite{fawz12} and \cite{carl92,carl95}
for previous results pertaining to magnetic and acoustic waves.

The differences in the magnetic energy generation at photospheric
heights go hand-in-hand with the decrease of stellar activity for
stars like $\beta$~Hyi relative to the Sun.  Observationally, it has
been found that there is a notably reduced Ca~II and Mg~II emission in
the line core fluxes, as well as evidence for cyclic chromospheric
activity; see, e.g., \cite{drav93b} and \cite{bucc08}.  Previous
studies about $\beta$~Hyi's upper atmosphere, notably the stellar
corona and wind, have been given by, e.g., \cite{drav93c}, \cite{gued98},
and \cite{pizz00}.
Based on those analyses, it is expected that the modes of energy propagation
and dissipation taken into account in this study, although nonmagnetic processes
are expected to dominate in evolved stars, are insufficient to
account for the vast variety of dynamic features in those regions, implying
the presence of other processes.  For moderately evolved stars, there is
also evidence pointing to the significance of 3-D structures, associated
with large-scale turbulence, as pointed out by \cite{judg93} and others.


\section*{Acknowledgments}

This work has been supported in part by the Faculty of Engineering,
Izmir University of Economics as well as the Department of Physics,
University of Texas at Arlington.  Furthermore, we are grateful to M. Sosebee
(UTA, Dep. of Physics) for his assistance with computer graphics.

\section*{Funding}

The authors do not report any funding.

\section*{Informed Consent}

All authors reviewed the manuscript and agreed to its content.

\section*{Conflict of Interest}

The authors declare that they have no competing interests as defined by Springer, or other interests that might be perceived to influence the results and/or discussion reported in this paper.

\section*{Data Availability}

The data that support the findings of this study are available from D.E.F. upon reasonable request.

\section*{Author Contribution}

D.E.F. pursued the calculations and prepared the figures.  M.C. has been the main contributor to the text.


\end{document}